\documentclass[conference]{IEEEtran}
\IEEEoverridecommandlockouts
\usepackage{cite}
\usepackage{amsmath,amssymb,amsfonts}
\usepackage{algorithmic}
\usepackage{graphicx}
\usepackage{textcomp}
\usepackage{xcolor}
\usepackage{tikz}
\usepackage{braket}
\usepackage{subcaption}

\def\beq{\begin{equation}}
\def\eeq{\end{equation}}
\def\beqn{\begin{equation*}}
\def\eeqn{\end{equation*}}
\def\bearn{\begin{eqnarray*}}
\def\eearn{\end{eqnarray*}}
\def\bear{\begin{eqnarray}}
\def\eear{\end{eqnarray}}
\def\barr{\begin{array}}
\def\earr{\end{array}}

\def\P{{ P}}

\def\LF{{\sf LIFO }}
\def\FF{{\sf FIFO }}
\def\LP{{\sf LIFO-PO}}
\def\LPm{{\text{\sf LIFO-PO}}}

\def\FP{{\sf FIFO-PO}}

\def\wsm{\prec^w}

\newcommand{\R}{\mathbb{R}}
\newcommand{\E}{\mathbb{E}}

\newtheorem{theorem}{\textbf{Theorem}}

\newtheorem{definition}{Definition}

\def \f{f}

\def\BibTeX{{\rm B\kern-.05em{\sc i\kern-.025em b}\kern-.08em
    T\kern-.1667em\lower.7ex\hbox{E}\kern-.125emX}}
\begin{document}

\title{Scheduling Quantum Teleportation with \\Noisy Memories\\
\thanks{This research was supported in part by the NSF grant CNS-1955834, NSF-ERC Center for Quantum Networks grant EEC-1941583 and by the National Science Foundation to the Computing Research Association for the CIFellows 2020 Program.}
}

\author{\IEEEauthorblockN{Aparimit Chandra$^{1}$, Wenhan~Dai$^{1,2}$, ~and Don~Towsley$^{1}$}
\\ $^1$College of Information and Computer Science, University of Massachusetts Amherst \\$^2$
Quantum Photonics Laboratory, Massachusetts Institute of Technology
\\[0.1em]
Email: aparimitchan@umass.edu, whdai@cs.umass.edu, and towsley@cs.umass.edu
}


\maketitle

\begin{abstract}
Quantum teleportation channels can overcome the effects of photonic loss, a major challenge in the implementation of a quantum network over fiber. Teleportation channels are created by distributing an entangled state between two nodes which is a probabilistic process requiring classical communication. This causes critical delays that can cause information loss as quantum data suffers from decoherence when stored in memory. In this work, we quantify the effect of decoherence on fidelity at a node in a quantum network due to the storage of qubits in noisy memory platforms. We model the memory platform as a buffer that stores incoming qubits waiting for the creation of a teleportation channel. Memory platforms are parameterized with decoherence rate and buffer size, in addition to the order in which the incoming qubits are served. We show that fidelity at a node is a linear sum of terms, exponentially decaying with time, where the decay rate depends on the decoherence rate of the memory platform. This allows us to utilize Laplace Transforms to derive efficiently computable functions of average fidelity with respect to the load, buffer size, and decoherence rate of the memory platform. We prove that serving qubits in a Last In First Out order with pushout for buffer overflow management is optimal in terms of average fidelity. Lastly, we apply this framework to model a single repeater node to calculate the average fidelity of the teleportation channels created by this repeater assuming perfect gate operations.
\end{abstract}


\section{Introduction}
Quantum networks face many problems inherently different from those in classical networks as qubits different from bits. One of these problems arises from the (in)famous no-cloning theorem. In addition Quantum networks implemented over fiber suffer from exponential photonic loss with respect to fiber length. Classical networks overcome similar loss by using signal amplification. Unfortunately, the no cloning theorem bars the use of signal amplification, which means quantum networks need to find another solution to the problem of loss. Quantum teleportation allows us to transfer quantum information between two spatially separated parties using a distributed entangled state and classical communication without having to transfer the physical entity carrying that information across the network making it invulnerable to loss. Another important property of quantum teleportation is that it allows for secure communication and is a central part of quantum key distribution. 

Quantum teleportation is enabled through the use of entangled quantum states, the most common example of which is an Einstein–Podolsky–Rosen (EPR) states or Bell Pairs. Therefore, a key job for quantum networking devices like repeaters, switches etc, is to distribute these EPR pairs between two nodes so that quantum information can be shared between them, creating a teleportation channel. Many protocols exist for the generation and distribution of EPR pairs but all of them are probabilistic processes that can fail because of imperfections in physical operations like gate errors, signal loss in fiber etc. These probabilistic failures naturally give rise to many optimisation, control and design problems in quantum networking devices, which is a field of active study. There have been many recent results regarding the modeling and analysis of entanglement distribution rates for nodes in quantum networks but a major assumption in most of these is the presence of noiseless memories. This is an issue as most noisy intermediate-scale quantum (NISQ) era quantum memory platforms cause fidelity loss on any qubit stored in them due to decoherence.

When a qubit arrives at a node requesting teleportation, the node must await the generation of an EPR pair between itself and the destination which causes delays. While the EPR pair is being generated, The request qubit has to be stored in a noisy memory platform. This means that even if the EPR pair is perfect, the request will suffer some decoherence. The sensitivity of a memory platform to noise is parameterized by decoherence time or rate and fidelity decays exponentially with time. Quantifying the effects of this decoherence on the fidelity of the teleportation allows us to come up with specifications for memory platforms for different applications. Another important question is memory management. Given the fact that one can only store a finite amount of requests and EPR pairs, how should one schedule and serve requests to minimize decoherence and to deal with arrival of new requests when memory is full

In this paper, we quantify the fidelity loss for a node of a quantum network due to decoherence from memory and provide a way to derive the fidelity distribution or the average fidelity. We use dephasing noise characterized by a dephasing rate $\Gamma$ capturing noise due to storage in a memory platform and derive an expression of fidelity with respect to time spent in memory by a request. We model the teleportation as a queuing process where requests are generated according to a Poisson process with rate $\lambda$ and the EPR are also generated according to a poisson process with rate $\mu$. This allows us to calculate the wait times in memory through the use of simple continuous time Markov models.

The average fidelity of a node depends on load, dephasing rate, memory size, and serving discipline. Because the relation between fidelity and time is not linear, the order in which requests are served can affect the fidelity. Even when the average wait times of two service disciplines are the same, the average fidelities that they can be different. We consider both \FF and \LF disciplines with both finite and infinite buffer variants: First in First out with infinite buffer(\FF), Last in First out infinite buffer(\LF), FIFO with finite buffer and pushout (\FP), LIFO with finite Buffer(\LP). In the case of a finite buffer, pushout means that if a request arrives to a full buffer, the oldest request in the buffer is kicked out to make space for the incoming qubit. We consider pushout because intuitively it optimizes for fidelity as older requests i.e. qubits that have suffered the most decoherence are kicked out. We give a proof that LIFO PO is indeed the optimal discipline for optimizing fidelity. 

We consider a scenario where we have two memory platforms available to us, One for storing teleportation requests and one for caching EPR pairs. We model this as two competing queues where at least one is always empty. Lastly, we extend this model to show how this can be applied to calculate the average fidelity of the teleportation channel created by a single repeater chain. The novelty in our construction stems from its simplicity and flexibility for calculating fidelity distributions. It also considers the effects of scheduling disciplines which, to the authors knowledge, at the time of writing have not been considered in quantum networks. The flexibility of this model also allows for easy extensions to different noise models, probability distributions etc. This leads to natural future work in integrating elements from different works. We will go further into this in section \ref{future_work}.

\subsection{Related work}
As stated previously the analysis and modelling of quantum network devices is an active field of study. There have been a lot of studies on modelling switches and repeaters to analyse and design protocols\cite{DRT}, \cite{VGNT2021}, but these studies focus Entanglement generation capacity regions not on fidelity. \cite{BCE20} focuses on fidelity of EPR pairs generated by repeater chains of different lengths but it does not account for a continuous stream of requests so it can be seen as deriving a more accurate distribution for the EPR generation distribution for a node at the beginning of a repeater chain. \cite{VSW22} analyses fidelity loss from wait times in memory using queues and is a very flexible model as it also abstracts away hardware implementations and protocols into tunable parameters, but it focuses mostly on local network on a quantum processor where as the model presented in this paper can be extended to model repeater nodes as well as other quantum nodes. They also don't consider different service disciplines considering only a \FF queue.

\section{System model}
In this section, we formally define the process we are modeling. We will define the parameters that govern our physical process how a memory platform in a quantum network node will behave. 

Consider a node in a quantum network. This node is constantly receiving information that it must process and then teleport to another node. We assume this node receives information as pure state qubits arriving according to a Poisson process with rate parameter $\lambda$. Any time this node receives a qubit, it must send it to some other node. It does so by generating an EPR pair between itself and the destination. We assume whenever this EPR pair is generated, it is initialized as a pure state and has fidelity 1. This may be a strong assumption, but we discuss later how it can be relaxed. 

Distribution of EPR pairs between two nodes is a stochastic process \cite{DPW20} where the probability of successful EPR pair generation depends on the distribution protocol and physical implementation of the EPR pair generating platform. If we consider a discrete time model, the number of time steps required to generate an EPR pair is characterized by a geometric distribution. If we consider the time for one trial to be very small, and the probability of successfully generating an EPR pair also to be small, then we can approximate the EPR generation process by a Poisson process where the time taken to generate an EPR pair is sampled from an exponential distribution with mean $\mu$. We define the "load" on a node as ${\lambda}/{\mu}$. We can then model the occupancy (number of stored qubits to be teleported and EPR pairs) of the memory platform as a continuous time Markov process (CTMC), and derive steady state distributions or Laplace transforms for the time a qubit spends in memory (wait time). We can then utilize memory error models to obtain statistical descriptions of the final fidelity of the teleported qubits.

\subsection{Memory Model}
A request qubit is stored in some noisy memory when waiting for an EPR pair. If more requests arrive while one is already in memory, a queue is formed. As stated previously, this allows us to model the memory platform as a CTMC allowing for the calculation of wait time distributions. To quantify the effects of decoherence, we need a continuous time noise model that captures information loss. We choose the dephasing or the phase damping model\cite{NC10} represented by the operator $\varepsilon(\rho)$, where $\rho$ is some density matrix of a one qubit system. It is mathematically defined as
\beq\label{error_model}
\varepsilon \left(\begin{bmatrix} \rho_{00}& \rho_{01}\\ \rho_{10}& \rho_{11} \end{bmatrix} \right)
= 
\begin{bmatrix} \rho_{00}& e^{-\Gamma t}\rho_{01}\\ e^{-\Gamma t}\rho_{10}& \rho_{11} \end{bmatrix} 
\eeq
where $\rho_{ij}$ is the $ij$th entry of the density matrix $\rho$, $\Gamma$ is a constant rate at which dephasing occurs in a given environment, and $t$ is time elapsed. Dephasing noise is the most common noise associated with memories and the results of this paper can be extended to account for any error model as long as it can be expressed as a liner sum of exponentially decaying terms.

\subsection{Fidelity loss of a single qubit}
Fidelity of some density matrix $\rho$ to some pure state $\ket{\psi}$ is given by the formula
\beq \label{fid}
F(\ket{\psi}\bra{\psi}, \rho) = \text{tr}(\ket{\psi}\bra{\psi} \rho).
\eeq
We can use this with \eqref{error_model}\ to calculate the fidelity of a single qubit $\ket{\psi} = \alpha\ket{0} + \beta\ket{1}$ after spending time $t$ in memory. We get the formula
\beq  \label{eq:Fid_1_qubit}
F(t) =  |\alpha|^4 + 2e^{-\Gamma t}|\alpha|^2|\beta|^2 + |\beta|^4.
\eeq
Note that the fidelity depends on the initial state of the pure qubit, i.e., $\alpha$ and $\beta$ influence the fidelity loss experienced by that qubit. We will need the inverse of this function to transform the distribution of wait times to the distribution for fidelity.
\beq  \label{eq:Fid_1_qubit_inv}
F^{-1}(t) =  \Gamma^{-1} (\ln(2|\alpha|^2|\beta|^2 - \ln(f - |\alpha|^4 - |\beta|^4)).
\eeq

\subsubsection{Fidelity loss in a Bell pair}
The effects of dephasing on the fidelity of Bell pair is well studied and is given by
\beq \label{eq:f_t1_T2}
F(t) = \frac{1+e^{-2\Gamma t}}{2} 
\eeq
where $t$ is time spent in the system and $\Gamma$ is again the dephasing rate of the memory \cite{MATN15}.

Dephasing causes the Bell state to turn into a mixture of a Bell state and maximally mixed 2-qubit state $I/4$ which can be written in terms of its fidelity to the Bell state as a non maximally entangled Bell state:
\beqn \label{eq:werner}
\rho_w = \frac{1-F}{3}I+\frac{4F -1}{3}\ket{\Phi^+}\bra{\Phi^+}.
\eeqn
Here $F$ is the fidelity of $\rho_w$ with respect to the Bell pair $\ket{\Phi^+}$. This is precisely the $F$ that decays in \eqref{eq:f_t1_T2}

\subsubsection{Fidelity loss experienced by a qubit due to teleportation by a non maximally entangled state}
Teleportation using a maximally entangled Bell pair results in perfect teleportation and no information is lost. This is rarely the case in practice so we take a look at how teleportation using a non maximally entangled Bell state state acts as a linear map on the input state. We consider a Werner state $\rho_w$ as the teleportation resource and use it to teleport $\rho(t)$ which is the density matrix for some request qubit 
that has spent $t$ time in memory. This allows us to represent the effect of teleportation on $\rho(t)$ as a linear map \cite{AFY}:
\beq \label{teleportation_operator}
\Lambda_{T}(\rho(t)) = \sum^{1}_{i, j = 0} \bra{\phi_{ij}}\rho_w\ket{\phi_{ij}}\cdot U_{ij}\rho(t)U_{ij}^{\dagger}
\eeq
where $\ket{\phi_{ij}}$ are Bell states, $\Lambda_{T}(\cdot)$ is the standard teleportation algorithm represented as a linear transformation and
$$U_{00} = I, U_{01} = \sigma_x, U_{10} = \sigma_z, U_{11} = i\sigma_y.$$
If $F$ is the fidelity of the Werner state $\rho_w$ with respect to the target Bell state  $\Phi^+$, then
\begin{align*}
\Lambda_{T}(\rho(t)) &= F \rho(t) + \frac{1-F}{3}\sigma_x\rho(t)\sigma_x^\dagger\\ &+ \frac{1-F}{3} \sigma_z\rho(t)\sigma_z^\dagger+ \frac{1-F}{3} i\sigma_y\rho(t)(i\sigma_y)^\dagger.
\end{align*}
 This equation can be further simplified to get an equation for the fidelity of a qubit being teleported by a non maximally entangled Bell state state both suffering dephasing errors for times $t_1$ and $t_2$ respectively. Therefore the final fidelity of the teleported qubit is
\begin{align}\nonumber
&\mathrm{tr}\big(\rho(0)\Lambda_{T}\rho(t_1)\big)\\ \nonumber
& = \frac{1+e^{-2\Gamma t_2}}{2} (|\alpha|^4 + |\beta|^4 +2e^{-\Gamma {t_1}}|\alpha|^2|\beta|^2)\\ \nonumber
& + \frac{1-e^{-2\Gamma t_2}}{6} (4e^{-\Gamma {t_1}}|\alpha|^2|\beta|^2)\\
& + \frac{1-e^{-2\Gamma t_2}}{6} (|\alpha|^4 + |\beta|^4 - e^{-\Gamma {t_1}}((\alpha^*\beta)^2 - (\beta^*\alpha)^2))\label{eq:fidelity_master}
\end{align}


In the considered model, either the EPR pairs or the request qubits have to be stored in memory so if $t_1 > 0$ then $t_2 = 0$ and vice versa. Therefore, we further simplify \eqref{eq:fidelity_master} in these cases.
If $t_1 = 0$, the error in teleportation is only due to dephasing suffered by the EPR pair, the formula simplifies to
\beqn \label{eq:F_w}
F_2(t) = \frac{3 + c_1}{6} + \frac{3 - c_1}{6} e^{-2 \Gamma t}, \quad t\ge 0
\eeqn
where $c_1 = 1 + 2(|{\alpha}|^2|{\beta}|^2 - (\alpha^*\beta)^2 - (\beta^*\alpha)^2)$. 
When $t_2 = 0$, we get the equation in \eqref{eq:Fid_1_qubit}:
\beqn
F_1(t) = c_2 + c_3e^{-\Gamma t}, \quad t\ge 0
\eeqn
where $c_2 = |{\alpha}|^4 + |{\beta}|^4$ and $c_3 = 2|{\alpha}|^2|{\beta}|^2$.
The key observation is that the fidelity is a linear sum of terms exponentially decaying with time. If time is a random variable and its Laplace transform with parameter $s$, denoted by $T^*(s)$ is known, we obtain the equation
\beq \label{eq:E[f]}
\E[F_i] = c_i + c_j\E[e^{-\Gamma_i T}] = c_i + c_j T^*(\Gamma_i).
\eeq
This approach of using the Laplace transform to get the moments for the fidelity is useful as in the processes we consider, it is easy to obtain closed form solutions of the Laplace transforms for the wait times than the actual distributions. This will be especially useful when we consider models with finite memory.
In these equations $c_i$ and $c_j$ are decided by the input qubit being teleported. In this paper we will use $\ket{+}$ as the example input qubit and
\beq \label{eq:F_+_simp}
F_1({t}) = \frac{1}{2} + \frac{1}{2}e^{-\Gamma_1 t}, \quad t\ge 0
\eeq
\beq \label{eq:F_+_w_simp}
F_2({t}) = \frac{2}{3} + \frac{1}{3} e^{-2 \Gamma_2 t}, \quad t\ge 0.
\eeq
These are the simplified error models being considered in this paper.
All of the aforementioned functions are monotonic scalar functions of the fidelity in terms of wait time of a request. We can transform the random variable for the wait time into the random variable for the fidelity. This is convenient as many wait time distributions have analogues in classical literature as we will see in the upcoming sections.

\section{Double Queue Model}
\begin{figure}
    \centering
\begin{tikzpicture}[scale=0.11] \label{fig:ctmcdq}
\tikzstyle{every node}+=[inner sep=0pt]
\draw [black] (4.4,-19.9) circle (3);
\draw (4.4,-19.9) node[scale=0.8] {$-B_e$};
\draw [black] (22.5,-19.9) circle (3);
\draw (22.5,-19.9) node {$-2$};
\draw [black] (40.6,-19.9) circle (3);
\draw (40.6,-19.9) node {$0$};
\draw [black] (48.9,-19.9) circle (3);
\draw (48.9,-19.9) node {$1$};
\draw [black] (13.4,-19.9) circle (3);
\draw (13.4,-19.9) node {$....$};
\draw [black] (31.3,-19.9) circle (3);
\draw (31.3,-19.9) node {$-1$};
\draw [black] (74.2,-19.9) circle (3);
\draw (74.2,-19.9) node[scale=0.8] {$B_r$};
\draw [black] (57.5,-19.9) circle (3);
\draw (57.5,-19.9) node {$2$};
\draw [black] (64.9,-19.9) circle (3);
\draw (64.9,-19.9) node {$....$};
\draw [black] (22.574,-16.954) arc (-200.09637:-339.90363:4.606);
\fill [black] (31.23,-16.95) -- (31.42,-16.03) -- (30.48,-16.37);
\draw (26.9,-13.43) node [above] {$\lambda_r$};
\draw [black] (31.317,-16.947) arc (-197.97018:-342.02982:4.871);
\fill [black] (40.58,-16.95) -- (40.81,-16.03) -- (39.86,-16.34);
\draw (35.95,-13.08) node [above] {$\lambda_r$};
\draw [black] (40.636,-16.958) arc (-200.29891:-339.70109:4.386);
\fill [black] (48.86,-16.96) -- (49.06,-16.03) -- (48.12,-16.38);
\draw (44.75,-13.59) node [above] {$\lambda_r$};
\draw [black] (13.551,-16.954) arc (-201.1594:-338.8406:4.717);
\fill [black] (22.35,-16.95) -- (22.53,-16.03) -- (21.59,-16.39);
\draw (17.95,-13.44) node [above] {$\lambda_r$};
\draw [black] (4.681,-16.966) arc (-204.07639:-335.92361:4.621);
\fill [black] (13.12,-16.97) -- (13.25,-16.03) -- (12.34,-16.44);
\draw (8.9,-13.73) node [above] {$\lambda_r$};
\draw [black] (48.956,-16.955) arc (-200.11224:-339.88776:4.519);
\fill [black] (57.44,-16.96) -- (57.64,-16.03) -- (56.7,-16.38);
\draw (53.2,-13.49) node [above] {$\lambda_r$};
\draw [black] (57.186,-16.982) arc (-194.57897:-345.42103:4.148);
\fill [black] (65.21,-16.98) -- (65.5,-16.08) -- (64.53,-16.33);
\draw (61.2,-13.38) node [above] {$\lambda_r$};
\draw [black] (65.149,-16.96) arc (-202.83905:-337.16095:4.775);
\fill [black] (73.95,-16.96) -- (74.1,-16.03) -- (73.18,-16.42);
\draw (69.55,-13.54) node [above] {$\lambda_r$};
\draw [black] (73.768,-22.818) arc (-26.63183:-153.36817:4.719);
\fill [black] (65.33,-22.82) -- (65.24,-23.76) -- (66.14,-23.31);
\draw (69.55,-25.92) node [below] {$\lambda_e$};
\draw [black] (65.127,-22.825) arc (-16.54566:-163.45434:4.096);
\fill [black] (57.27,-22.82) -- (57.02,-23.73) -- (57.98,-23.45);
\draw (61.2,-26.25) node [below] {$\lambda_e$};
\draw [black] (57.024,-22.803) arc (-28.97374:-151.02626:4.371);
\fill [black] (49.38,-22.8) -- (49.33,-23.74) -- (50.2,-23.26);
\draw (53.2,-25.56) node [below] {$\lambda_e$};
\draw [black] (48.917,-22.842) arc (-19.14458:-160.85542:4.411);
\fill [black] (40.58,-22.84) -- (40.37,-23.76) -- (41.32,-23.43);
\draw (44.75,-26.31) node [below] {$\lambda_e$};
\draw [black] (40.503,-22.851) arc (-19.66174:-160.33826:4.835);
\fill [black] (31.4,-22.85) -- (31.2,-23.77) -- (32.14,-23.44);
\draw (35.95,-26.56) node [below] {$\lambda_e$};
\draw [black] (31.203,-22.845) arc (-20.58332:-159.41668:4.596);
\fill [black] (22.6,-22.85) -- (22.41,-23.77) -- (23.35,-23.42);
\draw (26.9,-26.33) node [below] {$\lambda_e$};
\draw [black] (22.422,-22.849) arc (-19.62353:-160.37647:4.747);
\fill [black] (13.48,-22.85) -- (13.28,-23.77) -- (14.22,-23.43);
\draw (17.95,-26.5) node [below] {$\lambda_e$};
\draw [black] (13.379,-22.85) arc (-18.5988:-161.4012:4.726);
\fill [black] (4.42,-22.85) -- (4.2,-23.77) -- (5.15,-23.45);
\draw (8.9,-26.57) node [below] {$\lambda_e$};
\end{tikzpicture}
\caption{Markov Chain formulation of Double Queue model with buffer size $B_e$ for EPR pairs and $B_r$ for request qubits}
\label{fig:CTMC}
\end{figure}
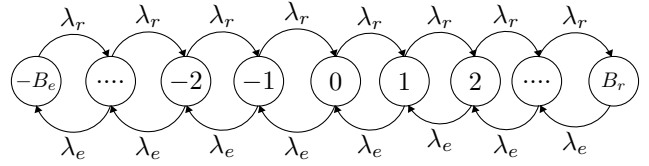

In this section, we consider a node that has a memory platform available for storing multiple EPR pairs which it generates according to a Poisson process with rate $\lambda_e$. Teleportation requests arrive accordig to a Poisson process with rate $\lambda_r$. We assume gate operations are instantaneous as the time taken to perform gate operations is orders of magnitude smaller than the time taken to generate an EPR pair. This process can be modeled as two competing queues where the service rate for one queue is the request rate for another. The memory platform for the request qubits can store $B_r$ qubits and the platform for EPR pairs can store $B_e$ qubits. We can model this as a CTMC with the state being the number of request qubits in the system denoted as $N$. We consider a surplus of EPR pairs as having negative requests making $-B_e\leq N \leq B_r$. This gives us the process presented in Fig. \ref{fig:CTMC}
From its Markov Chain formulation, 
\beqn
\pi_n = \mathbb{P}[{N} = n] = \pi_{-B_e} \rho^{n+B_e}
\eeqn
Since $\pi_{-B_e} \sum_{i = 0}^{B_e +B_r} \rho^i = 1$, 
\beq 
\pi_n = \frac{1-\rho}{1-\rho^{B_e + B_r +1}} \rho^{n + B_e}.
\eeq

Let $p_e$ and $p_r$ denote probability an arriving EPR pair and request gets placed in a buffer, respectively. This means the EPR pair or the request has no counterpart to pair up with. Then
\begin{align*}
   p_e = \sum_{n = -\text{B}_e}^0 \pi_n \text{ and }
p_r = \sum^{\text{B}_r}_{n = 0} \pi_n.
\end{align*}

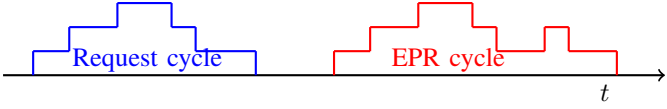
\begin{figure}[t]
\centerline{
		\begin{tikzpicture}[thick, scale = 0.8]
        \draw[->,thick] (1,0) -- (12,0) ;
        \draw[blue,thick] (1.5,0) -- (1.5,0.4) ;
        \draw[blue,thick] (1.5,0.4) -- (2.1,0.4) ;
        \draw[blue,thick] (2.1,0.4) -- (2.1,0.8) ;
        \draw[blue,thick] (2.1,0.8) -- (2.9,0.8) ;
        \draw[blue,thick] (2.9,0.8) -- (2.9,1.2) ;
        \draw[blue,thick] (2.9,1.2) -- (3.8,1.2) ;
        \draw[blue,thick] (3.8,1.2) -- (3.8,0.8) ;
        \draw[blue,thick] (3.8,0.8) -- (4.2,0.8) ;
        \draw[blue,thick] (4.2,0.8) -- (4.2,0.4) ;
        \draw[blue,thick] (4.2,0.4) -- (5.2,0.4) ;
        \draw[blue,thick] (5.2,0.4) -- (5.2,0) ;
        \node at (3.4,.25) {\color{blue}Request cycle};
        \draw[red,thick] (6.5,0) -- (6.5,0.4) ;
        \draw[red,thick] (6.5,0.4) -- (7.1,0.4) ;
        \draw[red,thick] (7.1,0.4) -- (7.1,0.8) ;
        \draw[red,thick] (7.1,0.8) -- (7.9,0.8) ;
        \draw[red,thick] (7.9,0.8) -- (7.9,1.2) ;
        \draw[red,thick] (7.9,1.2) -- (8.8,1.2) ;
        \draw[red,thick] (8.8,1.2) -- (8.8,0.8) ;
        \draw[red,thick] (8.8,0.8) -- (9.2,0.8) ;
        \draw[red,thick] (9.2,0.8) -- (9.2,0.4) ;
        \draw[red,thick] (9.2,0.4) -- (10,0.4) ;
        \draw[red,thick] (10,0.4) -- (10,0.8) ;
        \draw[red,thick] (10,0.8) -- (10.4,0.8) ;
        \draw[red,thick] (10.4,0.8) -- (10.4,0.4) ;
        \draw[red,thick] (10.4,0.4) -- (11.2,0.4) ;
        \draw[red,thick] (11.2,0.4) -- (11.2,0) ;
        \node at (8.4,.25) {\color{red}EPR cycle};
        \node at (11,-0.3) {$t$};
		\end{tikzpicture}
}
        \caption{Typical teleportation behavior.} \label{fig:service-cycles}
\end{figure}

Let $P_{s, r}$ and $P_{s, e}$ be the probabilities that a request qubit is teleported and an EPR pair is used, respectively. Let $i \in \{e, r\}$. 
Define: $P_{s, i} = \mathbb{P}[\text{an arrival of type $i$ gets served}]$. Then
\beq \label{eq:prob_served_PO_disc}
P_{s,i} = \frac{\sum^{B_i - 1}_{j = 0}\rho^j}{\sum^{B_i}_{j = 0}\rho^j} = \frac{1-\rho^{B_i}}{1-\rho^{B_i+1}}
\eeq
We see from the Markov chain formulation that this system alternates between two phases as shown in figure \ref{fig:service-cycles}. In phase 1 request qubits are stored in memory waiting for  EPR pairs and  suffer decoherence during the wait.  In phase 2 EPR pairs queue up in memory and wait for requests to arrive that they can teleport.  The system alternates between these two phases so fidelity loss suffered by the teleported qubit is sometimes due to decoherence of request qubit and sometimes due to the decoherence EPR pair but never both due to the assumption that gate operation are instantaneous and error free and initially both the request qubit and EPR pair are pure states. This means we can individually analyse the fidelity loss for each phase and then derive a joint distribution by conditioning on the phase.

Let us take a closer look at the process happening in one phase. If we restrict ourselves to one phase, the memory platform behaves like a standard finite buffer M/M/1. Now we know from queuing theory that different orders of service for buffered requests and EPR pairs lead to different wait time distributions. If we have have the wait time distributions, we can easily derive fidelity distributions using a Jacobian transformation. When it is too complex to explicitly derive the wait time distribution and it is usually straightforward to calculate the Laplace transform for the wait time and use it to calculate average fidelity.

Let $\f_{W_{i}}(t)$ be the probability density function (pdf) for the wait time incurred by a random request during phase $i = 1,2$. The qubit fidelity distribution will depend on whether the request qubit or the EPR pair incurred the wait (phases 1 and 2). In our case, we will use \eqref{eq:F_+_simp} or \eqref{eq:F_+_w_simp} depending on what type of qubit we are considering.
\beqn{}
\f_{F_{i}}(x) = f_W\big(F_i^{-1}(x)\big)  \left| \frac{d}{dy} \big(F_i^{-1}(x)\big) \right|.
\eeqn{}
We also define $W^*_{i}(s) =\E[e^{-st}]$, i.e., the Laplace transform of the wait time during phase $i = 1,2$. We use \eqref{eq:E[f]} to calculate $\E[F_i]$ given $W^*_{i}$.

In the next section, we give explicit expressions for $\f_{F_{i}}(t)$ or $W^*_{i}(s)$ for four memory platforms differentiated by the type of memory management or service discipline used. Meanwhile, assuming we have descriptions for the wait time of the two individual queues, we can get the joint probability distribution of fidelity of served requests accounting for both phases by adding the conditional distributions of fidelity of a served request which waited in a particular queue and normalizing it. We get the expression
\beq \label{eq:phase_conditioning_equation}
\f_F(x) = \frac{\lambda_e p_e \P_{s,e}\f_{F_{e}}(x) + \lambda_r p_r \P_{s,r} \f_{F_{r}}(x)}{\lambda_e p_e\P_{s,e} + \lambda_r p_r\P_{s,r}}.
\eeq
Therefore,
\beq \label{eq:Ephase_conditioning_equation}
\E[F] = \frac{\lambda_e p_e \P_{s,e}\E[F_e] + \lambda_r p_r \P_{s,r} \E[F_r]}{\lambda_e p_e\P_{s,e} + \lambda_r p_r\P_{s,r}}.
\eeq
This expression is very flexible as it allows us to calculate the average fidelity for double queue models even if the memory platforms have different decoherence rates, buffer sizes, control etc in a modular manner

\section{Single Queue Models}
In this section, we take a closer look at the phases mentioned in the previous section by focusing on a  single memory platform for the incoming ``request'' and has no memory for the ``service''. Applied to the double queue model we need to be careful whether the ``request'' is an actual request qubit or an EPR pair as they will flip depending on which phase we are in. Once the request arrives in memory, it waits for service. As stated before inter arrival and service times are sampled from exponential distributions making this identical to an M/M/1 queue. We will now take a look at five different kinds of queues each modelling a different kind of memory control for the platform

\subsection{Infinite buffer model. FIFO}
The buffer size is infinite in this system and no incoming request is blocked. The buffer serves the qubits in a FIFO discipline. We know from literature \cite{Z} that 
\[\f_W(t) = (\lambda_e-\lambda_r)e^{-(\lambda_e-\lambda_r)t}, \quad t\ge 0\]
where $\lambda_r$ is the arrival rate and $\lambda_e$ is the service rate. We can transform this into the probability density function for the fidelity using the formula derived in \eqref{eq:Fid_1_qubit} as the function is scalar and monotonic to get
\beq \label{eq:fid_dist_FIFO}
\begin{split}
\f_F(x) &= \frac{(\lambda_e - \lambda_r)}{\exp({\ln{|\alpha|^4 + |\beta|^4} - \ln{x - 2|\alpha|^2|\beta|^2}})^{(\lambda_e - \lambda_r)/\Gamma}}\\&\quad \cdot|\frac{\Gamma^{-1}}{(|\alpha|^4 + |\beta|^4 - x)}|
\end{split}
\eeq
The Laplace transform for $f_W(t)$ is
\beq \label{eq:fid_dist_FIFO}
W^*(\Gamma) = \frac{\lambda_e - \lambda_r}{\lambda_e - \lambda_r -\Gamma}.
\eeq
One key thing about this model is that it is only valid if $\lambda_r < \lambda_e$ as the queue will just keep growing and there will be no steady state. If we are optimising for fidelity, we need to minimise wait times. Serving in LIFO discipline intuitively makes more sense especially at higher loads, as we want to prioritise younger requests.
\subsection{Infinite Buffer LIFO}
This is very similar to the previous model except that the derivation of the wait time distribution is different. In this we will consider a system requests are served in a LIFO order i.e. the buffer is a stack. the buffer has infinite length. We will use the following parameters to model the system. The busy period of a queue is defined as the time period measured between the instant a request arrives to an empty system until another request leaves behind an empty buffer. For an M/M/1 queue, the distribution of the busy period is given by
\beq \label{eq:busyPeriod_mm1}
\f_B(t) = \frac{1}{t\sqrt{\rho}}e^{-(\lambda_r+\lambda_e)t}I_1(2t\sqrt{\lambda_r\lambda_e})
\eeq
where $\rho = \lambda_r/\lambda_e$ is the load. Since a new request is always placed in front of the buffer, the  wait time distribution is the same as the busy period distribution \cite{Z}, i.e.,
\beq\label{eq:pdf_wait_time_lifo}
\f_W(t) = \f_B(t).
\eeq
With the inverse of the function of fidelity with respect to time
\beq
g^{-1}(f) =  \Gamma^{-1} (\ln(2|\alpha|^2|\beta|^2 - \ln(f - |\alpha|^4 - |\beta|^4))
\eeq
we can now transform the pdf of wait time into the pdf for the fidelity 
\beq\label{pdf_lifo}
\f_F(x) = \f_B(g^{-1}(x))|\frac{\Gamma^{-1}}{(|\alpha|^4 + |\beta|^4 - x)}|.
\eeq

We also know the Laplace of the busy period and by extension of the wait time is:
\beq
W^*(\Gamma)= \frac{1}{2 \lambda_r}(\lambda_r + \lambda_e + \Gamma - \sqrt{(\lambda_r + \lambda_e + \Gamma)^2 - 4 \lambda_r \lambda_e})
\eeq


\subsection{Finite buffer FIFO with pushout}
In this section we consider a system in which the incoming requests are stored in a queue with a finite maximum buffer capacity $B$. If a request arrives when the queue is full, the oldest request in the queue is discarded and the incoming request is stored in the queue. This makes the probability of service for a requesting \FP dependent not only on its position in the queue ($k$) but also on the number of qubits behind it ($j$) as it might get pushed out otherwise. We need to define a new probability, $W_r(j, k, t)$ which is the probability that a request in position $k$ with $j$ requests behind it is served and its remaining wait time is $t$.

We know from results that the Laplace transform $W^*(j,k,s) = \int^\infty_0 e^{-s t}W(j, k, t) dt$ can be described by the following set of recursive equations\cite{DH86}
\begin{align*}
W^*(j, 0, s ) &= 1 \\
W^*(B-1, 1, s)  &= \frac{\lambda_e}{\lambda_e + \lambda_r + s}\\
W^*(j, k, s)  &= \frac{\lambda_r}{\lambda_r + \lambda_e + s}W^*(j+1, k, s)\\
&+\frac{\lambda_e}{\lambda_r  + \lambda_e + s} W^*_r(j, k-1, s)\\
W^*(B -k, k, s)  &= \frac{\lambda_r}{\lambda_r + \lambda_e + s} W^*(B - k + 1, k - 1, s)\\
&+\frac{\lambda_e}{\lambda_r + \lambda_e + s} W^*(B -k, k-1, s).
\end{align*}
These equations can now be used to define
\beqn
W^*(s) = \Big[\sum^{B_r} _{j = 1} W^*(j, k, s)\Big] + W^*(N, 0, s)
\eeqn
Since this is a joint probaility of service and wait time. We need to turn this into a conditional probability of waiting time given the request will be served which we can get by normalizing with probability of a random request getting service $P_s$ Therefore the Laplace would be
\beqn
\E[e^{-\Gamma W}] = \frac{W^*(s)}{P_s}
\eeqn
This can be used to calculate $\E[F]$. Next we consider \LP.

\subsection{Finite buffer LIFO with pushout}
The main difference between this section and the previous one is that incoming request qubits are stored in a stack instead of a queue. We still discard the oldest qubit if a request arrives when it is full and the incoming request is put on the top of the stack. Unfortunately it is difficult to work directly with the wait time pdf. Instead we work with the  Laplace Transform. Unlike the previous model the wait time only depends upon the its position in the queue $k$. Let $W(k, t)$ denote the probability density that a request in buffer position $k$ gets served eventually and its wait time will be $t$. Assuming the head of the queue starts at 1, From classical results,
\beq
\begin{split}
W^*(0, s) &= 1, W^*(B_i+1, s) = 0,\\
W^*(k,s) &= \frac{\lambda_i}{\lambda_e+\lambda_r+s}W^*(k+1, s)\\ &+ \frac{\lambda_{i'}}{\lambda_e+\lambda_r+s}W^*(k-1, s),\\
W^*(B, s) &= \frac{\lambda_{i'}}{\lambda_e+\lambda_r+s}W(B-1, s),\\
W^*(1, s) &= \frac{\lambda_{i}}{\lambda_e+\lambda_r+s}W^*(2, s)\\ &+  \frac{\lambda_{i'}}{\lambda_e+\lambda_r+s}W^*(B-1, s).
\end{split}
\eeq
They can be solved to produce
\beq \label{eq:LT_lifopo}
W^*(k, s) = \frac{r_1(s)^k r_2(s)^B - r_2(s)^k r_1(s)^B}{r_2(s)^B - r_2(s)^k}
\eeq
where
$$r_{1, 2}(s) = \frac{(\lambda_e+\lambda_r+s) \pm \sqrt{(\lambda_e+\lambda_r+s)^2 - 4\lambda_e\lambda_r}}{2\lambda_i}.$$
Since a new request arriving at LIFO always goes in the first position,
\beqn
W^*(s) = W^*(1, s).
\eeqn
We need to normalize this Laplace transform as in the previous section with $P_s$ to get
\beqn
\E[e^{-sW}] = \frac{W^*(1,s)}{P_s}.
\eeqn



\section{Optimality of LIFO-PO}\label{sec:opt}
We have analyzed several memory management and service disciplines.  This raises the question as to which performs best.  In this secction we answer this question by establishing that out of a large class of work conserving disciplines, \LP~is optimal in that it maximizes final average fidelity of a teleported qubit.
 This result should not come as a surprise as it is  well known that, out of the class of work conserving non-preemptive policies $\Pi'$, LIFO maximizes $\E[f(W^\pi )]$ for any convex function $f$ where $W^\pi$ is the sojourn time under policy $\pi \in \Pi'$ for an infinite buffer G/G/1 queue \cite{SS87,LNT95}.    


Let $\pi$ denote a policy that assigns requests to EPR pairs and determines what teleportation qubits and EPR pairs to remove from the respective buffers to avoid overflow. We first observe that there is no benefit to removing a qubit from a buffer before it is full; hence we only consider polices that remove qubits at the time overflow occurs.  Second, we restrict ourselves to work conserving policies; those that always teleport qubits whenever possible.  Let $\Pi$ denote the set of such double buffer policies.  We introduce \LP, which always assigns the youngest qubit to be teleported to a newly created EPR pair or the youngest EPR pair to a newly made teleportation request, and always throws out the oldest qubit from the buffer when it is about to overflow. A formal definition of this policy is given in the Appendix.
Henceforth we refer to \LP~as $\gamma$.
\begin{theorem} \label{th:optimality}
Out of the class of policies $\Pi$, \LP~maximizes average fidelity,
\[\E [F^\pi] \le \E [F^\LPm], \quad \forall \pi \in \Pi. \] 
where $F_\pi$ is the teleportation fidelity under $\pi$.
\end{theorem}
{\em Proof sketch.} A complete proof is found in the appendix.  Here we provide a sketch of the proof.  The system can be decomposed into two single buffer subsystems, one for teleportation requests, the other for EPR pairs.  Let $F_\mathrm{e}^\pi$ and $F_\mathrm{r}^\pi$ denote the fidelity for EPR pairs and teleportation requests respectively.  We show that $\E [F_\mathrm{e}^\pi]$ and $\E[F_\mathrm{r}^\pi]$ are maximized when $\pi = \LPm$. As $\E [F^\pi]$ is a weighted average of $\E [F_\mathrm{e}^\pi]$ and $\E[F_\mathrm{r}^\pi]$, this establishes the theorem.

Focusing on the request buffer, we condition on the first $n$ departures of qubits from the request buffer, either due to successful teleportation or removal due to overflow. Let $w^\pi = (w^\pi_1, \ldots ,w^\pi_n)$ denote the wait times of these requests. Because request qubits can be removed from the buffer without service, we will assign wait times of infinity to those requests. Let $m$ denote the number of these removed qubits.
Our proof that $\gamma$ is optimal is based on establishing the following majorization result between $w^\pi$ and $w^\gamma$, $\pi \in \Pi$, $\pi \ne \gamma$, $w^{\pi} \wsm w^{\gamma}$. Here $\wsm$ is defined as follows.
\begin{definition}
Let $x,y\in\R_+^{n-m}\times\{\infty\}^m$; $y$ weakly supermajorizes $x$ written $x\wsm y$ iff
\[
  \sum_{i=1}^k x_{(i)} \ge \sum_{i=1}^k y_{(i)}, \quad k=1,\ldots , n-m.
\] 
where $x_{(i)}$ (resp. $y_{(i)}$) correspond to the components of $x$ ($y$) in increasing order.\\
\end{definition}
This is useful in our context because of the following property of $\wsm$, 
 \beq \label{eq:cxequiv}
 \sum_{i=1}^{n-m} \phi(x_{(i)}) \le \sum_{i=1}^{n-m} \phi(y_{(i)})
 \eeq
 for any continuous decreasing convex function $\phi$.
 
The proof that $w^{\pi} \wsm w^{\gamma}$ is straightforward and consists of transforming $\pi$ into $\gamma$ by taking each non-\LP~decision and replacing it by an \LP~decision such that the weak majorization is propagated until the resulting policy is \LP. Property (\ref{eq:cxequiv}) can now be applied with $\phi() = F()$ where $F()$ is given in (\ref{eq:F_+_simp}), (\ref{eq:F_+_w_simp}), $n$ allowed to go to infinity, and the conditioning on arrival and departure times removed yielding $\E [F(W_r^\pi )] \le \E[F(W_r^\LP)]$. The EPR buffer is handled in a similar manner. 

\section{Results}\label{sec:results}
\begin{figure}[t]
\centerline{\includegraphics[scale = 0.5]{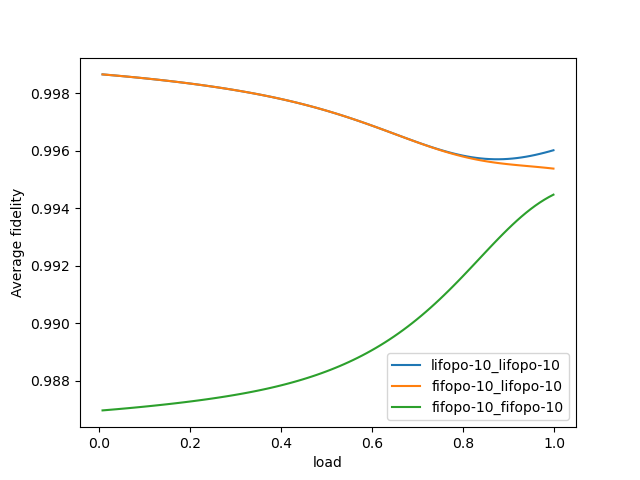}}
\caption{Plot for average fidelity vs. load. $\Gamma = 0.01$, $\lambda_e = 5$ and $\lambda_r \in (0,5)$, therefore, load $\in [0,1]$}
\label{fig:dqplots}
\end{figure}
\begin{figure}[t]
\centerline{\includegraphics[scale = 0.5]{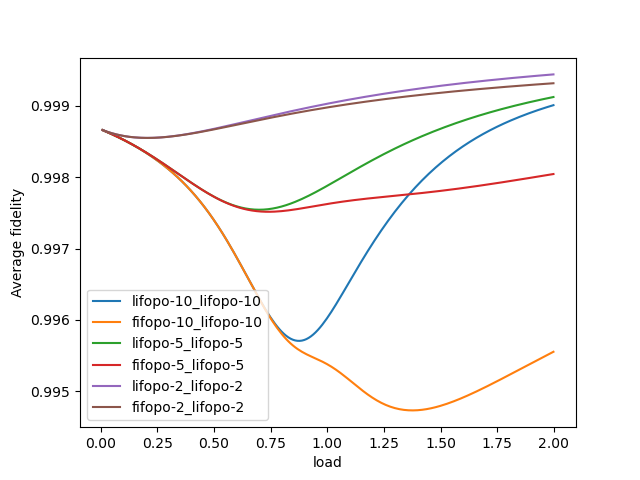}}
\caption{Plot for average fidelity vs. load comparing different buffer sizes. $\Gamma = 0.01$, $\lambda_e = 5$ and $\lambda_r \in (0,10)$, therefore, load $\in [0,2]$}
\label{fig:dqplot2}
\end{figure}

\begin{figure}[t]
\centerline{\includegraphics[scale = 0.5]{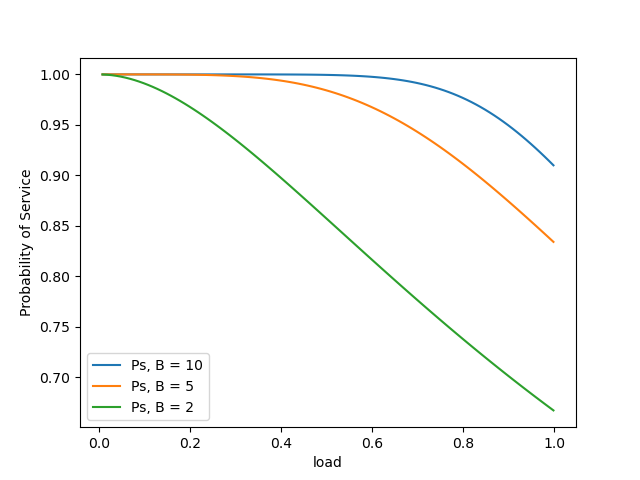}}
\caption{Plot for Probability a random request reciever service vs. load for a \LP-\LP~queue.}
\label{fig:ps}
\end{figure}
\begin{figure}[t]
\centerline{\includegraphics[scale = 0.5]{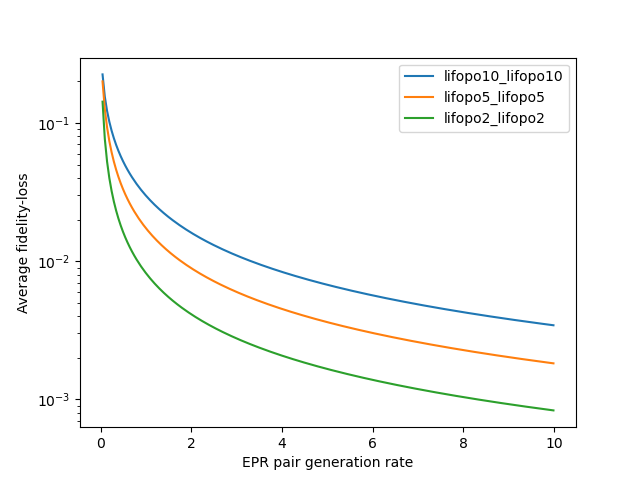}}
\caption{Plot for average fidelity loss vs. load for a single repeater chain for different buffer sizes following \LP-\LP. $\Gamma = 0.01$.}
\label{fig:dqr}
\end{figure}

\begin{figure*}[t]
\centerline{\includegraphics[scale = 0.4]{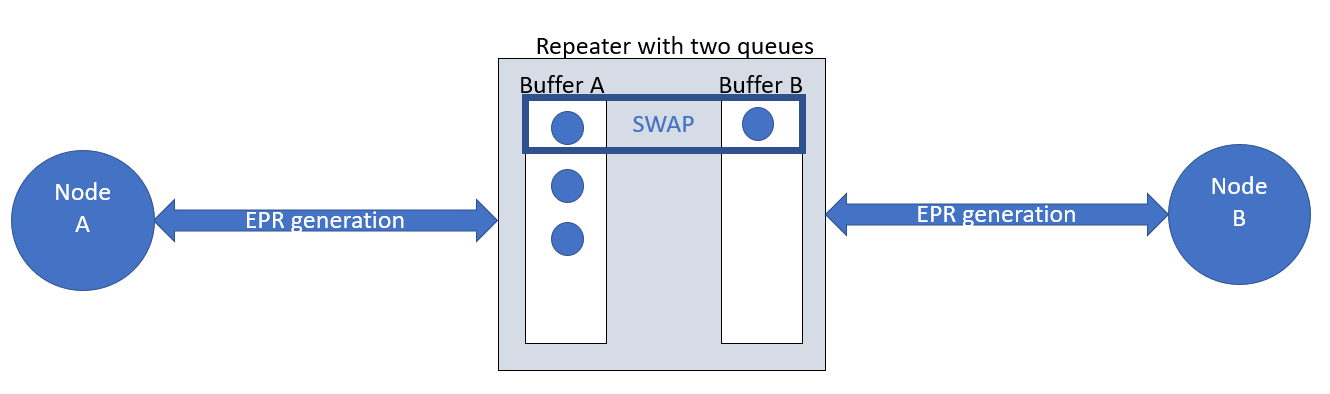}}
\caption{A single repeater between two nodes, It has two buffers to store EPR pairs and only one can non empty at the same time. If the repeater has a qubit from an EPR pair in buffer A, and an EPR pair is generated between it and B, it performs a swap and discards the qubits.}
\label{fig:repeater}
\end{figure*}

It was proven in Section \ref{sec:opt} that \LP~maximizes average fidelity. We can visualize this in Figure \ref{fig:dqplots}. We plot the average fidelity of a teleported request with respect to load for different service disciplines, \FP-\LP, \LP-\LP~and \FP-\FP. Buffer sizes are set to $B = 10$, dephasing rate $\Gamma = 0.01$, the EPR generation rate is $\lambda_e = 5$, the teleportation request rate $\lambda_r$ between zero and ten. We observe that \LP~-\LP~outperforms \LP~-\FP~and \FP~-\FP. The reason for the lower performance of \FP-\FP~at low loads is explained by the fact that the EPR pairs are being queued up waiting for requests but since they have to be used in order of creation, the requests are served by stale EPR pairs rather then fresh ones as is the case of systems that use \LP~for the EPR pair. Of course \LP-\LP~performs the best which is consistent with Theorem \ref{th:optimality}. Another thing to note is the increasing nature of \FP-\FP. Since we have a pushout mechanism for the oldest qubit in the queue, increasing the load means a greater chance of older requests being kicked out. In the case of \LP-\LP, we observe that as load approaches one, the fidelity stops decreasing and starts increasing. This because EPR pairs are served more quickly and have incurred less decoherence. 

In Figure \ref{fig:dqplot2} we explore the performance of \LP-\LP~and \LP-\FP~as a function of load for three different buffer sizes of 2, 5, 10.  Here we allow load to vary from zero to two ($\lambda_r$ varies from zero to 10).  In all cases average fidelity first decreases and then increases.In the case of a buffer size of 10, the minimum occurs close to a load of one.  This behavior should not come as a surprise as the time qubits spend in either buffer is the same at load one.  The reason that minimum average fidelity does not occur at load one is due to the fact that request qubits and EPR pairs decohere at different rates and the asymmetry becomes more pronounced as buffer size decreases. Last average fidelity decreases with buffer size because increasing the size allows qubits more time to decohere before use. 

Figure \ref{fig:ps} examines the behavior of probability of service as a function of load for the three different \LP-\LP~buffer sizes 


\subsection{Application to  one repeater node}

In this section, we apply the double queue model to a repeater between two nodes A and B as show in Figure \ref{fig:repeater}. This repeater is constantly generating EPR pairs between A and B by generating EPR pairs between itself and A, and itself and B, and then performing entanglement swaps. Entanglement swapping is a form of teleportation  one qubit of an EPR pair `a' using EPR pair `b'. This ``swaps'' the entanglement as now one of the qubits of the resource EPR pair `b' has been entangled with the non-teleported qubit of EPR pair `a' and the original entanglements has been destroyed. This repeater behaves like two queues with one queue modelling the memory of EPR pairs between repeater and node A and the other between the repeater and node B. 

Our teleportation model requires the following modification, namely request qubits suffer the same type of decoherence as EPR pairs. However, we get around this by assuming that the repeater always teleports using the newest generated EPR pair as a resource. This ensures that the fidelity of a resource state is always 1. Since we are always teleporting with a maximally entangled state, the fidelity of the final EPR pair generated between A and B is equivalent to the fidelity loss suffered by the EPR pair which was waiting in memory as that qubit will be perfectly teleported. The fidelity of a bell pair dephasing with time is given by \eqref{eq:f_t1_T2} which is the fidelity function we use for both the queues.

As modeled before, EPR pairs are generated according to a Poisson process and the time between consecutive EPR generations between the repeater and node $x$ is sampled from an exponential distribution with mean $\mu_x$, $x = A,B$. Now assuming the router is equidistant between node A and node B and it uses identical technologies for generating the EPR pair, $\mu_A = \mu_B$. To keep it consistent, we keep the decoherence rate, $\Gamma = 0.01$. We plot the average infidelity defined as $1 - \E[F] $with respect to $\mu$ on a log scale in Figure \ref{fig:dqr}. Average infidelity decreases with increasing rate because when the rate is low, one queue receives a pair but since the other queue has a low rate, the arrived pair has to wait a long time before it has a counterpart for service. We also see larger infidelity in larger buffers but again this comes at the cost of a greater chance of rejection as seen in Figure \ref{fig:ps}.


\section{Summary}
In this paper, we have modeled and quantified the effects of decoherence in a Teleportation Process, we model memory platforms in networks as queues and utilize queuing theory to calculate how much time a request has to wait for teleportation. We then map these waiting times to fidelity loss due to dephasing. This allows us to derive efficiently computable functions to calculate average fidelity of the qubits teleported by a node. We consider a case where there are two queues to model caching of EPR pairs and provide a framework to extend results from classical queuing theory about single buffer queues to the double buffer systems. We quantify how serving disciplines can significantly affect teleportation fidelities in NISQ era devices and calculate average fidelities for different disciplines. We prove the optimality of \LP-\LP~ for serving teleportation requests and compare it to other disciplines. We analyze the effects of buffer sizes and give a comparison of their Service probabilities. Lastly we apply this framework to analyze the average transportation fidelity of a quantum repeater between two nodes and see how different buffer sizes compare in terms of fidelity and service completion probability.

\subsection{Future Work} \label{future_work}
There are many open questions and directions this work can take. A most natural extension is to account for mixed states as requests. One can achieve this by modifying \eqref{eq:fidelity_master} and use the fidelity formula for comparing two mixed states instead of assuming a state is pure. Another direction would be to use more accurate distribution models for the EPR pair generation as in \cite{BCE20} and apply this model to longer repeater chains. Another natural extension would be to model a constant timeout policy so that if a request has been in the queue for longer than some time $C$ so that we can guarantee a minimum fidelity for the teleported information.

\appendix

{\bf Proof of Theorem \ref{th:optimality}.} We focus separately on the two buffers and focus on the amount of time qubits to be teleported and EPR pairs are allowed to decohere waiting to be matched up with each other.  Henceforth we focus on the request buffer and only on requests that arrive when no EPR pair is stored in the EPR buffer. We focus on the arrivals and departures of $n$ requests under policy $\pi \in \Pi$.  Let $a_1,\ldots , a_n$ and $d_1, \ldots , d_n$ denote the arrival and departure times for these requests.  

Here a departure corresponds either to a pairing with a newly creation of an EPR pair followed by a successful teleportation or removal due to buffer overflow. Let  $m\le n$ denote the qubits removed from the buffer.   \LP~satisfies the following properties: 
\begin{itemize}
    \item There exists no pair of requests $j,k$ that are served such that  $a_k < a_j < d_k < d_j$,
    \item there exists no pair of requests $j,k$ where $k$ is served and $j$ is discarded such that $a_k < a_j < d_k$,
    \item there exists no pair of requests $j,K$ that are discarded such that $a_k < a_j < d_j < d_k$
\end{itemize}  
  Let $w^\pi = (w^\pi_1, \ldots ,w^\pi_n)$ denote the wait times of these teleportation requests. Because requests can be removed from the buffer without service, we will assign wait times of infinity to those requests. 
Our proof that $\gamma$ is optimal is based on showing $w^\pi \wsm w^\gamma$, $\pi \in \Pi$. 
Note that the standard definition \cite{MOA11} corresponds to the case $m=0$. We introduce an operator $T_{ij}$, called the "$T$-transform", as follows.  Let $x\in \R_+^n$;
\[ 
T_{ij} = \lambda I + (1-\lambda )Q_{ij}
\]
where $I$ is the identity operator, $Q_{ij}$ is an operator that permutes the $i$-th and $j$-th components of $x$ and $0\le \lambda \le 1$.  In other words,
\begin{align*}
T_{ij}x &= (x_1,\ldots ,x_{i-1},\lambda x_i +(1-\lambda)x_j,x_{i+1}, \ldots, \\
& x_{j-1},(1-\lambda)x_i + \lambda x_j,x_{j+1}, \ldots , x_n)
\end{align*}
It is easily shown that $T_{ij}x \wsm x$ provided $x_i,x_j < \infty$.  Note that $x \wsm Q_{ij}x$ ($\lambda = 0$). Last, define the function $S_j(x)$ as
\[
 S_j(x) = (x_1,\ldots , x_{j-1},\alpha x_j,x_{j+1}, \ldots , x_n)
 \]
 with $0\leq\alpha \leq 1$.  Then $x \wsm S_j(x)$.
 
 Consider the system with $n$ requests arriving at times $a_1,\ldots , a_n$ and depart at times $d_1, \ldots , d_n$. 

 We transform $\pi$ to $\gamma$ through a sequence of steps that creates a sequence of policies $\pi_0 = \pi, \pi_1, \pi_2, \ldots , \pi_h = \gamma$ such that $w^{\pi_l} \wsm w^{\pi_{l+1}}$, $l=0, \ldots h-1$.
 
 Assume $\pi_l$ violates the above property of {\LP}. There are three cases depending on whether the two requests are served, one is served and the other removed or both removed.
 \begin{enumerate}
     \item {\bf Both are served.}  Request $k$ is served before a younger request $j$, $a_k < a_j < d_k < d_j$  (we omit dependence on $\pi_l$). We construct $\pi_{l+1}$ from $\pi_l$ by switching the order in which $j$ and $k$ are served. The wait times for requests $j$ and $k$ under $\pi_l$ are $w^{\pi_l}_j = d_j - a_j$ and $w^{\pi_l}_k = d_k - a_k$ and under $\pi_{l+1}$ are $w^{\pi_{l+1}}_j = d_k - a_j$ and $w^{\pi_{l+1}}_k = d_j - a_k$. Here $w^{\pi_l}$ and $w^{\pi_{l+1}}$ satisfy
     \[
       w^{\pi_l} = T_{jk}w^{\pi_{l+1}} 
     \]
     with 
     \[
        \lambda = \frac{a_j - a_k} {(a_j-a_k)+(d_j - d_k)}.
     \]
     Hence $w^{\pi_l} \wsm w^{\pi_{l+1}}.$ See Figure \ref{fig:case1}.
     \item {\bf One request is served.}  Request $k$ is served while a younger request is discarded, $a_k < a_j < d_k$. We switch the order in which these two requests are handled resulting in the servicing of $j$ at time $d_k$ and removal of $k$ at time $d_j$.   Then $w^{\pi_l} $ and $w^{\pi_{l+1}}$ satisfy
     \[
       w^{\pi_{l+1}} = S(Q_{jk}w^{\pi_l} )
     \]
     with $\alpha = (d_k - a_j)/(d_k - a_k).$
     Hence $w^{\pi_l} \wsm w^{\pi_{l+1}}.$ See Figure \ref{fig:case2}.
     \item {\bf Both are removed.} A younger request $j$ is removed from the buffer before an older job $k$ under $\pi_l$, $a_k < a_j < d_j < d_k$.  We switch the order of the removals under $\pi_{l+1}$.  This does not affect wait times and $w^{\pi_l} \wsm w^{\pi_{l+1}}.$ See Figure \ref{fig:case3}.
 \end{enumerate}
This procedure is repeated until the properties of \LP~are satisfied and, consequently $w^{\pi} \wsm w^{\gamma}$.

We fixed the arrival and service times. Remove the conditioning on them and let $W^{\pi}(n)$ denote the wait time of a randomly chosen request from the first $n$ requests that are served. From the above majorization result and the equivalence (\ref{eq:cxequiv}), we conclude that $\E [\phi (W^{\LP}(n)] \ge \E [\phi (W^{\pi}(n))]$ for every convex decreasing function $\phi$. Moreover if the limits $W^{\LP} = \lim_{n\rightarrow \infty} W^{\LP}(n)$ and $W^{\pi} = \lim_{n\rightarrow \infty} W^{\pi}(n)$ exist, then $\E [\phi (W^{\LP}] \ge \E [W^{\pi}]$.
\begin{figure}[ht]
\centerline{
		\begin{tikzpicture}[thick]
        \draw[thick] (1,0) -- (6.5,0) node[pos=0,left]{$\pi_l$};
        \draw[->,thick] (1.5,-0.5) -- (1.5,0) node[pos=0,right]{$a_k$};
        \draw[->,thick] (3,-0.5) -- (3,0) node[pos=0,right]{$a_j$};
        \draw[->,thick] (5,0) -- (5,0.5) node[pos=1,right]{$k$};
        \node at (5,-0.5) {$d_k$};
        \draw[->,thick] (6,0) -- (6,0.5)  node[pos=1,right]{$j$};
        \node at (6,-0.5) {$d_j$};
        \node (1) at (1.4,0.25){};
        \node (2) at (3.25,0.25) {$w_k^{\pi_l}$};
        \node (3) at (5.1,0.25){};
        \path[->]
        (2) edge (1) [thick] ;
        \path[->] (2) edge (3) [thick] ;
        \node (4) at (2.9,0.75){};
        \node (5) at (4.5,0.75) {$w_j^{\pi_l}$};
        \node (6) at (6.1,0.75){};
        \path[->]
        (5) edge (4) [thick] ;
        \path[->] (5) edge (6) [thick] ;
        \draw[thick] (1,-2) -- (6.5,-2) node[pos=0,left]{$\pi_{l+1}$}; 
        \draw[->,thick] (1.5,-2.5) -- (1.5,-2) node[pos=0,right]{$a_k$};
        \draw[->,thick] (3,-2.5) -- (3,-2) node[pos=0,right]{$a_j$};
        \draw[->,thick] (5,-2) -- (5,-1.5) node[pos=1,right]{$j$};
        \node at (5,-2.5) {$d_k$};
        \draw[->,thick] (6,-2) -- (6,-1.5) node[pos=1,right]{$k$};
        \node at (6,-2.5) {$d_j$};
        \node (7) at (1.4,-1.25){};
        \node (8) at (3.75,-1.25) {$w_k^{\pi_{l+1}}$};
        \node (9) at (5.1,-1.75){};
        \node (10) at (2.9,-1.75){};
        \node (11) at (4,-1.75) {$w_j^{\pi_{l+1}}$};
        \node (12) at (6.1,-1.25){};
        \path[->]
        (11) edge (10) [thick] ;
        \path[->] (11) edge (9) [thick] ;
        \path[->]
        (8) edge (7) [thick] ;
        \path[->] (8) edge (12) [thick] ;
		\end{tikzpicture}
}
        \caption{Case 1.} \label{fig:case1}
\end{figure}
\begin{figure}
\centerline{
		\begin{tikzpicture}[thick]
        \draw[thick] (1,0) -- (6.5,0) node[pos=0,left]{$\pi_l$};
        \draw[->,thick] (1.5,-0.5) -- (1.5,0) node[pos=0,right]{$a_k$};
        \draw[->,thick] (3,-0.5) -- (3,0) node[pos=0,right]{$a_j$};
        \draw[->,thick] (5,0) -- (5,0.5) node[pos=1,right]{$k$};
        \node at (5,-0.5) {$d_k$};
        \node at (6,0.5) {$j$};
        \node at (6,0) {\textcolor{red}{\bf{\sf X}}};
        \node at (6,-0.5) {$d_j$};
        \node (1) at (1.4,0.4){};
        \node (2) at (3.25,0.4) {$w_k^{\pi_l}$};
        \node (3) at (5.1,0.4){};
        \path[->]
        (2) edge (1) [thick] ;
        \path[->] (2) edge (3) [thick] ;
        \draw[thick] (1,-2) -- (6.5,-2) node[pos=0,left]{$\pi_{l+1}$}; 
        \draw[->,thick] (1.5,-2.5) -- (1.5,-2) node[pos=0,right]{$a_k$};
        \draw[->,thick] (3,-2.5) -- (3,-2) node[pos=0,right]{$a_j$};
        \draw[->,thick] (5,-2) -- (5,-1.5) node[pos=1,right]{$j$};
        \node at (5,-2.5) {$d_k$};
        \node at (6,0-1.5) {$k$};
        \node at (6,-2) {\textcolor{red}{\bf{\sf X}}};
        \node at (6,-2.5) {$d_j$};
        \node (9) at (5.1,-1.6){};
        \node (10) at (2.9,-1.6){};
        \node (11) at (4,-1.6) {$w_j^{\pi_{l+1}}$};
        \path[->]
        (11) edge (10) [thick] ;
        \path[->] (11) edge (9) [thick] ;
		\end{tikzpicture}
}
		\caption{Case 2.} \label{fig:case2}
\end{figure}
\begin{figure}[t]
\centerline{
		\begin{tikzpicture}[thick]
        \draw[thick] (1,0) -- (6.5,0) node[pos=0,left]{$\pi_l$};
        \draw[->,thick] (1.5,-0.5) -- (1.5,0) node[pos=0,right]{$a_k$};
        \draw[->,thick] (3,-0.5) -- (3,0) node[pos=0,right]{$a_j$};
        \node at (5,0.5) {$j$};
        \node at (5,0) {\textcolor{red}{\bf{\sf X}}};
        \node at (5,-0.5) {$d_j$};
        \node at (6,0.5) {$k$};
        \node at (6,0) {\textcolor{red}{\bf{\sf X}}};
        \node at (6,-0.5) {$d_k$};
        \draw[thick] (1,-2) -- (6.5,-2) node[pos=0,left]{$\pi_{l+1}$}; 
        \draw[->,thick] (1.5,-2.5) -- (1.5,-2) node[pos=0,right]{$a_k$};
        \draw[->,thick] (3,-2.5) -- (3,-2) node[pos=0,right]{$a_j$};
        \node at (5,-1.5) {$k$};
        \node at (5,-2) {\textcolor{red}{\bf{\sf X}}};
        \node at (5,-2.5) {$d_j$};
        \node at (6,0-1.5) {$j$};
        \node at (6,-2) {\textcolor{red}{\bf{\sf X}}};
        \node at (6,-2.5) {$d_k$};
		\end{tikzpicture}
}
		\caption{Case 3.} \label{fig:case3}
\end{figure}

Returning to our teleportation system, under the assumption that requests and EPR pairs are generated according to Poisson processes, when placed in their respective buffers, they will exhibit
stationary wait time $W^\pi_r$ and $W^\pi_e$ respectively.  The respective qubits decohere at different rates $F_r(t)$ and $F_e(t)$ in the two memories according to (\ref{eq:F_+_simp}), (\ref{eq:F_+_w_simp}), As these decoherence functions are decreasing and convex, we conclude that  $\E [F_r (W^{\LP}(n)] \ge \E [F_r (W^{\pi}(n))]$ and $\E [F_e (W^{\LP}(n)] \ge \E [F_e (W^{\pi}(n))]$ where $F_r$ is. The expected teleportation fidelity for the entire system, $\E[F^\pi]$ is $\E[F^\pi] = q\E[F^\pi] + (1-q) \E[F^\pi]$ where  $q$ is the probability that a request qubit arrives to a system where no EPR qubits are available.  Finally, we conclude $\E[F^{\LP}] \ge \E[F^\pi]$ for all $\pi \in \Pi$.

\end{document}